\documentclass[aps,pra,reprint,groupedaddress,amsmath,amssymb,preprintnumbers,showpacs]{revtex4-1}
\usepackage{enumerate}
\usepackage{graphicx}
\usepackage{color}
\usepackage{float}
\usepackage{xspace}
\graphicspath{{./}}

\newcommand{\ket}[1]{\bigl| #1 \bigr>} % for Dirac bras
\newcommand{\bra}[1]{\bigl< #1 \bigr|} % for Dirac kets
\newcommand{\braket}[2]{\bigl< #1 \vphantom{#2} \bigr|
 \bigl. #2 \vphantom{#1} \bigr>} % for Dirac brackets
\newcommand{\abs}[1]{\left| #1 \right|} % for absolute value

\newcommand{\pdt}[3]{\frac{\partial^2 #1}{\partial #2 \partial #3}}

\newcommand{\mol}{${\text{H}_2}^+$\xspace}
\newcommand{\floqu}[2]{u^{#2}_{#1}\left(x,R\right)} % Floquet state
\newcommand{\floqel}[1]{\xi_{E_\text{e}}}
\newcommand{\floqnuc}[1]{\xi_{E_\text{N}}}

\newcommand{\floqcel}[1]{\xi^{P}_{{E_\text{e}}}\left(#1\right)}
\newcommand{\floqcnuc}[1]{\chi_{{E_\text{N}}}\left(#1\right)}
\newcommand{\coulombr}{F_{E_\text{e}}(x;R,\alpha_0)}
\newcommand{\coulombi}{G_{E_{\text{e}}}(x;R,\alpha_0)}
\newcommand{\ee}{E_\text{e}}
\newcommand{\en}{E_\text{N}}

\begin{document}
\title{Dissociative ionization of \mol using intense femtosecond XUV laser pulses}
\author{Lun \surname{Yue}}
\author{Lars Bojer \surname{Madsen}}
\affiliation{Department of Physics and Astronomy, Aarhus University, DK-8000 Aarhus C, Denmark}
\date{\today}

\begin{abstract}
The dissociative ionization of \mol interacting with intense, femtosecond extreme-ultraviolet laser pulses is investigated theoretically. This is done by numerical propagation of the time-dependent Schr\"{o}dinger equation for a colinear one-dimensional model of \mol, with electronic and nuclear motion treated exactly within the limitations of the model. The joint energy spectra (JESs) are extracted for the fragmented electron and nuclei by means of the time-dependent surface flux method. The dynamic interference effect, which was first observed in one-electron atomic systems, is in the present work observed for \mol, emerging as interference patterns in the JESs. The photoelectron spectrum and the nuclear energy spectrum are obtained by integration of the JESs. Without the JESs, the photoelectron spectrum itself is shown to be inadequate for the observation of the dynamic interference effect. The resulting JESs are analyzed in terms of two models. In one model the wave function is expanded in terms of the ``essential'' states of the system, consisting of the ground state and the continuum states, allowing for the interpretation of the main features in the JESs in terms of dynamic interference. In the second model the photoelectron spectra from fixed-nuclei calculations are used to reproduce some features of the JESs using simple reflection arguments. The range of validity of these models is discussed and it is shown that
 the consideration of the population of excited vibrational states is crucial for understanding the structures of the JESs.

\end{abstract}

%\pacs{33.80.Rv, 31.15.xv, 33.20.Xx}
\pacs{33.20.Xx,  82.50.Kx, 33.80.Eh}
% 31.15.xv : Molecular dynamics and other numerical methods (Scrinzi t-SURFF)
% 33.80.Rv : Multiphoton ionization and excitation to highly excited states (e.g., Rydberg states) (Leth PRL og CBM PRL)
% 33.20.Xx : Spectra induced by strong-field or attosecond laser irradiation (CBM PRL)
% 33.80.Eh 	Autoionization, photoionization, and photodetachment
% 82.50.Kx 	Processes caused by X-rays or γ-rays
%CEDERBAUM: 33.20.Xx, 41.60.Cr, 82.50.Kx
%33.20.Xx: Spectra induced by strong-field or attosecond laser irradiation 
%41.60.Cr: Free-electron lasers (see also 52.59.Rz Free-electron devices—in plasma physics)
%82.50.Kx: Processes caused by X-rays or γ-rays
%CHUAN: 32.80.Rm, 82.50.Kx
%32.80.Rm 	Multiphoton ionization and excitation to highly excited states
%82.50.Kx 	Processes caused by X-rays or γ-rays
\maketitle

\section{Introduction}
The rapid developments in laser technology have made it possible to produce extreme-ultraviolet (XUV) laser pulses in the femtosecond to subfemtosecond time domains \cite{Sansone11,Ackermann07,McNeil10}. These new light sources have led to a range of novel attosecond time-resolved probing techniques, such as attosecond streaking spectroscopy \cite{Kienberger02,Schultze10}, time-resolved inner-shell spectroscopy \cite{Drescher02}, attosecond transient absorption spectroscopy \cite{Wang10}, and attosecond interferometry \cite{Remetter06}. Due to new focusing techniques, XUV pulses with very high peak intensities in the femtosecond time domain have been achieved \cite{Mashiko04,Sorokin07}. 
Projects such as the pan-European ELI \cite{ELI} lead us to expect that even higher laser intensities will be realizable in the future.

There has recently been theoretical interest in the ionization of simple atomic systems using such intense XUV laser pulses \cite{Toyota07,Toyota08,Demekhin12,Demekhin13,Demekhin13b,Yu13a}. Intensity modulations in the photoelectron spectra (PESs) were observed and
explained as follows. Due to the external field, the field-dressed ground-state energy is shifted in time by the ac Stark energy shift, which follows the laser field intensity envelope.
There are two times during the pulse at which 
there is resonance with the same continuum energy, once in the rising part and the other in the falling part of the laser pulse \cite{Toyota07,Demekhin12}. The two electronic wave packets ionized at the two different times pick up different phases during the duration of the pulse and interfere in the continuum, resulting in the interference structure observed in the PESs. 
This ac-Stark-shift-induced effect is referred to as dynamic interference \cite{Demekhin12}. 

Following the prediction of dynamic interference in atomic systems interacting with strong XUV laser pulses, it is natural to ask the question whether the same effect occurs in molecules.
One work addressed this question by considering dissociative ionization of \mol in a one-dimensional (1D) model \cite{Yu13c}, where dynamic interference was reported to occur in the PES. In that work, the Born-Oppenheimer (BO) approximation was employed, and the wave function was expanded in terms of the BO electronic ground state and approximate continuum states represented by plane waves. Continuum-continuum couplings, which should play a role for the intense pulses considered, were neglected. In this work, we investigate dissociative ionization of \mol by direct numerical propagation of the time-dependent Schr\"{o}dinger equation (TDSE) for a reduced-dimensionality model of \mol, treating both the electronic and nuclear degrees of freedom exactly within the limitations of the reduced dimensions. Such models have been studied extensively in the literature and reproduce experimental results at least qualitatively for lasers in the visible and infrared regimes \cite{Kulander96,Steeg03,Weixing01}. Recently, the dynamic interference effect observed in a one-dimensional model of hydrogen was compared to the effect in the three-dimensional case, and the one-dimensional model was shown to be adequate for the qualitative description of dynamic interference \cite{Yu13a}. Here, we focus on the joint energy spectrum (JES), which provides the differential probability of measuring a given electronic and nuclear kinetic energy. For interaction with strong near-infrared light, the JES is known to contain much more physical information than the standard PES and nuclear energy spectrum (NES) \cite{Madsen12,Silva13,Wu13,Yue13}. The aim of the present work is to demonstrate that dynamic interference occurs in \mol, and that the JES is crucial for the detection of the effect. When the JES is integrated to obtain the PES and NES, the interference patterns will be severely or completely washed out, and important information will be lost.

Aside from the exact numerical propagation of the TDSE, we employ two models to analyze the JES. 
One model consists of making an ansatz for the wave function consisting of only the initial state and the final continuum states, making the BO approximation and the rotating-wave approximation (RWA) and neglecting continuum-continuum couplings. In situations where the ground-state amplitude and the laser envelope vary slowly compared to the dynamical phase associated  with the detuning of the energy of the combined electron-nuclei system with respect to the central carrier frequency, a closed analytical expression for the JES can be obtained. By evaluation of this expression  in the saddle-point approximation, the structures in the JESs can be linked to the dynamical interference effect. 
In the other model, we construct the JESs from fixed-nuclei TDSE calculations by applying simple reflection arguments. Both models fail to work for certain laser parameters, and the range of applicability and the reasons for the failure will be discussed.

Obtaining spectra from numerical simulations is a challenge.
Standard methods of obtaining PESs from numerical calculations include projection on plane waves \cite{Madsen07,Demekhin13b}, projection on scattering states \cite{Fernandez09,Madsen12}, and the usage of flux methods \cite{Feuerstein03a,Keller95,Chelkowski98,Tao12}. In the first method, huge simulation volumes are required, as it must be ensured that after the end of the pulse, the scattered parts of the wave packet are well separated from its bound part and not reflected from the box boundaries.
In the second method,
the simulation volume can be reduced somewhat compared to the first method due to the orthogonality between the scattering and bound states, and the projection can be performed immediately at the end of the pulse. However, the construction of scattering states is tedious and constitutes a numerical challenge in itself. For the flux methods, absorbers are placed at the boundaries of the simulation volume to remove the outgoing flux, and spectra are obtained by monitoring the flux going through surfaces placed 
at distances smaller than the absorber regions. In this way the simulation volume can be reduced significantly.
We calculate the JES of \mol by employing a flux method, the time-dependent surface flux (t-SURFF) method \cite{Yue13} (see Refs.~\cite{Tao12,Scrinzi12,Serov13,Karamatskou14} for application to single atoms), which reduces the numerical effort significantly compared to other methods \cite{Madsen12,Silva13}.

The paper is organized as follows. In Sec.~\ref{sec:theory}, the reduced-dimensionality model for ${\text{H}_2}^+$ is described, and the extraction of JES using the t-SURFF method is outlined. In Sec.~\ref{sec:num_res}, exact numerical results for the dissociative ionization of \mol are presented for different pulse parameters. In Sec.~\ref{sec:analysis}, the JES is analyzed in terms of two models. Section~\ref{sec:conclusion} concludes the work. Atomic units are used throughout, unless indicated otherwise.

\section{Theory}
\label{sec:theory}
\subsection{Model for ${\text{H}_2}^+$}
We consider a simplified model for ${\text{H}_2}^+$ with reduced dimensionality that includes only the dimension that is aligned with a linearly polarized laser pulse \cite{Kulander96,Steeg03,Weixing01}. Within this model, electronic and nuclear degrees of freedom are treated exactly. The center-of-mass motion of the molecule can be separated, such that the TDSE for the relative motion in the dipole approximation and velocity gauge reads 
\begin{equation}
  i\partial_t\ket{\Psi(t)}=H(t) \ket{\Psi(t)}
  \label{eq:TDSE}
\end{equation}
with the Hamiltonian
\begin{equation}
  \begin{aligned}
    H(t)
    =T_\text{e}+T_\text{N}+V_\text{eN}+V_\text{N}+V_\text{I}(t),
    \label{eq:Hamiltonian}
    \end{aligned}
\end{equation}
where  $\ket{\Psi(t)}$ in coordinate space depends on the internuclear distance $R$ and the electronic coordinate  $x$  measured with respect to the center of mass of the nuclei. The components of the Hamiltonian in Eq.~\eqref{eq:Hamiltonian} are $T_\text{e}=-(1/2\mu)\partial^2/\partial x^2$, $T_\text{N}=-(1/m_\text{p})\partial^2/\partial R^2$, $V_\text{eN}(x,R)=-1/\sqrt{{(x-R/2)^2+a(R)}}-1/\sqrt{(x+R/2)^2+a(R)}$, $V_\text{N}(R)=1/R$, and $V_\text{I}(t)=-i \beta A(t) \partial/\partial x $, where $m_\text{p}=1.836\times 10^3$ a.u. is the proton mass, $\mu=2m_\text{p}/(2m_\text{p}+1)$ is the reduced electron mass, $\beta = (m_\text{p}+1)/m_\text{p}$, and the softening parameter $a(R)$ for the Coulomb singularity is chosen to produce the exact three-dimensional $1s\sigma_g$ BO potential energy curve \cite{Madsen12,Yue13}.

We use vector potentials of the form 
\begin{equation}
  A(t)=A_0g(t)\cos(\omega t),
  \label{eq:vecpot}
\end{equation}
where $\omega$ is the carrier angular frequency and $A_0$ is the amplitude chosen such that $\omega^2A_0^2=I$, with $I$ the intensity.
The field envelope is taken to be of Gaussian form,
\begin{equation}
  g(t)=\exp\left(-4\ln 2\frac{t^2}{\tau^2}\right),
  \label{eq:las_env}
\end{equation}
where $\tau$ is the full width at half maximum of the field envelope.

Equation \eqref{eq:TDSE} is solved exactly on a two-dimensional spatial grid using the split-operator, fast Fourier transform method \cite{Feit82}, with a time-step of $\Delta t=0.005$ in the time-propagation. The grid size is defined by $\abs{x}\le 200$ and  $R\le 60$, with grid spacings $\Delta x=0.391$ and $\Delta R=0.059$. These parameters ensured converged results.

\subsection{Extraction of JES}
The JES gives the differential probability for observing a nuclear kinetic energy of $E_\text{N}=k^2/m_\text{p}$ and an electron with kinetic energy $E_\text{e}=p^2/2\mu$, and can be calculated using the expression 
\begin{equation}
  \begin{aligned}
    \pdt{P}{E_\text{e}}{E_\text{N}}=\sum_{\text{sgn}(p)}\frac{m_\text{p}\mu}{2\abs{p}k}\abs{b_{p,k}(T)}^2,
    \label{eq:DI-JES}
  \end{aligned}
\end{equation}
where the summation over $\text{sgn}(p)$ refers to the summation of $\pm p$ corresponding to the same $E_\text{e}$, and $T$ is a sufficiently large time after the end of the laser pulse when the flux corresponding to dissociative ionization has moved into the asymptotic regions. As the Gaussian pulses used have asymptotic tails that extend to infinity, we take $T$ to be the smallest time at which the JES has converged numerically.
The JES is obtained from the TDSE calculations by using the molecular t-SURFF method described in Ref.~\cite{Yue13}. In this method, the coordinate space is partitioned into spatial regions corresponding to different reaction channels. The amplitudes $b_{p,k}(T)$ in Eq.~\eqref{eq:DI-JES} are written as
\begin{equation}
  b_{p,k}(T)=\braket{\phi_p(T)\chi_k(T)}{\Psi_\text{DI}(T)},
  \label{eq:DI_bampl}
\end{equation}
where $\phi_p(x,T)$ and $\chi_k(R,T)$ are plane waves for the electron and nuclei, respectively, and $\Psi_\text{DI}(x,R,T)$ is the wave packet corresponding to dissociative ionization at large $x$ and $R$. We denote the positions of the electronic and nuclear boundaries of the spatial region corresponding to dissociative ionization as $x_\text{s}$ and $R_\text{s}$, respectively. Equation~\eqref{eq:DI_bampl} is then 
evaluated by the method of Ref.~\cite{Yue13}, where the amplitudes $b_{p,k}(T)$ are rewritten into time integrals over the flux going through surfaces placed at $x=x_\text{s}$ and $R=R_\text{s}$. 
Evaluation of the time integrals requires the usage of Volkov waves instead of the plane waves $\phi_p(x,T)$.
In the present calculations we choose $x_\text{s}=100$, $R_\text{s}=20$, and $T=4\tau+800$.

The t-SURFF method allows the use of absorbers to remove the outgoing flux, avoiding unphysical reflections at the boundaries of the simulation volume and thus significantly reducing the size of the simulation volume. We make use of complex absorbing potentials (CAPs) with the form \cite{Muga04}
\begin{equation}
  V_\text{CAP}(r)=
  \begin{cases}
  -i\eta \left(\abs{r}-r_\text{CAP}\right)^n  ,& |r| \geq r_\text{CAP}, \\
  0 & \text{elsewhere},
  \end{cases}
  \label{eq:CAP}
\end{equation}
with $r$ being either the electronic coordinate $x$ or the nuclear coordinate $R$. We choose $\eta_e=0.001$, $x_\text{CAP}=110$ and $n_e=2$ for the electronic CAP, and $\eta_\text{N}=0.001$, $R_\text{CAP}=25$, and $n_\text{N}=2$ for the nuclear CAP.

\section{Numerical results}
\label{sec:num_res}
\begin{figure}
  \centering
  \includegraphics[width=0.45\textwidth]{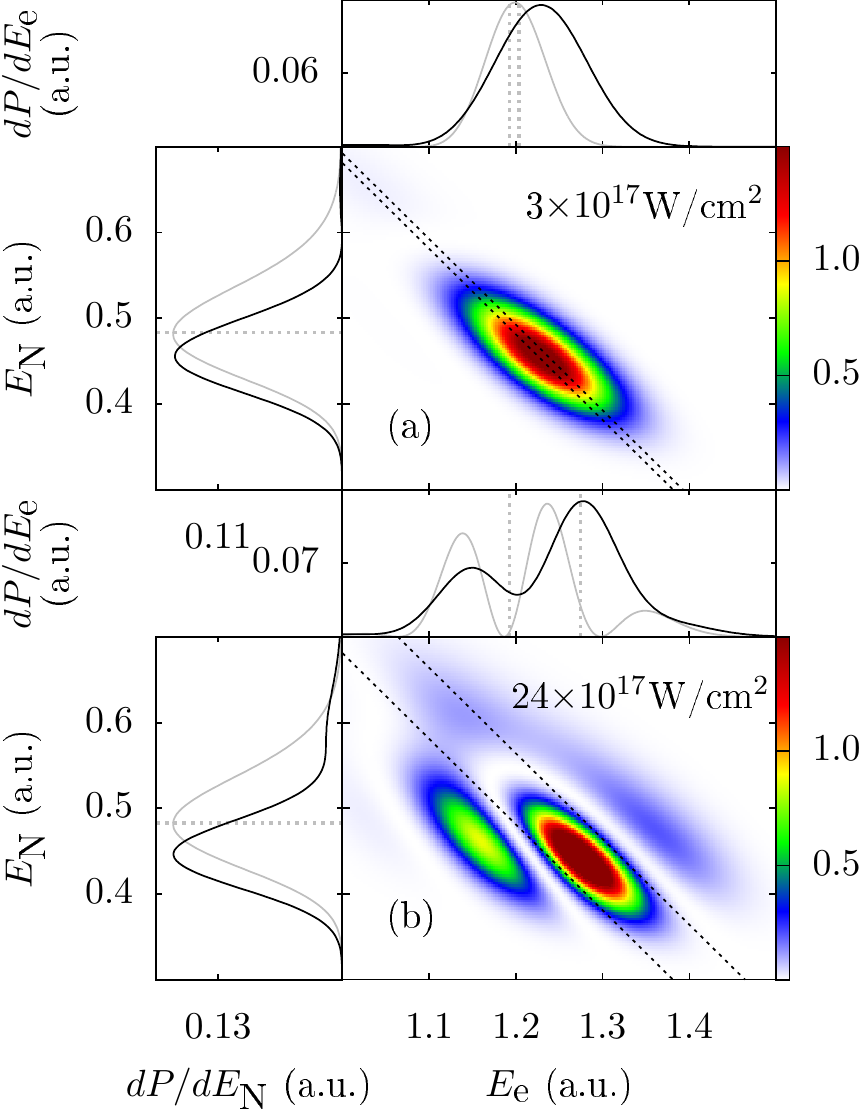}
\caption{(Color online) Spectra of \mol exposed to pulses with $\omega=2.278$, $\tau=1.1$ fs, and two different intensities $I$. Main panels show the JES [Eq.~\eqref{eq:DI-JES}]. The left and right dashed diagonal lines indicate the energy conservations $E_\text{e}+E_\text{N}=E_0+\omega$ and $E_\text{e}+E_\text{N}=E_0+\omega+\Delta$, respectively, with $\Delta$ the Stark shift [Eq.~\eqref{eq:tdse_stark}]. In the upper insets, thick, black curves show the PESs obtained by integration of the JESs, while the thin, gray curves show the PESs for nuclei fixed at the equilibrium distance $R_0=2.06$. The fixed-nuclei PES are scaled for better comparability with the moving nuclei PES. The dashed grey vertical lines show the energy conservations and Stark-shifted energies for the fixed-nuclei calculations. In the side insets, thick, black curves show the NESs obtained by integration of the JESs, while the thin, gray curves show the scaled-reflection-principle results [Eq.~\eqref{eq:tdse_reflection}]. The horizontal dashed gray lines indicate the position $E_\text{N}=1/R_0$.}
  \label{fig1}
\end{figure}

In order to investigate the JES, we first prepare \mol in its ground state $\ket{\Psi_0}$, obtained by imaginary propagation of Eq.~\eqref{eq:TDSE}. The ground-state energy is found to be $E_0=-0.597$. Convergence of all results is checked 
with respect to different box parameters, CAP parameters, and placement of t-SURFF surfaces.
The carrier frequency of the laser pulse is chosen as $\omega=2.278$, allowing for dissociative ionization by one-photon absorption, and the intensities are in the range from $I=3\times 10^{17}$ W/cm$^2$ to $I=24\times 10^{17}$ W/cm$^2$.
In this work, we are specifically interested in the first JES peak corresponding to one-photon absorption, located close along the diagonal line $E_\text{N}+E_\text{e}\approx E_0+\omega$ in the JES (see Fig.~\ref{fig1}). 
Note that although the pulses considered in this work are ultraintense, we are still in the nonrelativistic regime. 
Indeed, if the cycle-averaged quiver energy of a free electron (the ponderomotive energy) $U_\text{p}=A_0^2/4$ is much less than its rest energy $m_\text{e}c^2=137^2$, we are in the nonrelativistic regime \cite{Sarachik70,Mourou06,Boyd08}. This condition is certainly satisfied here, as even for the most intense pulse $I=24\times 10^{17}$ W/cm$^2$ the ponderomotive energy is only $U_\text{p}=3.29$.

Figure~\ref{fig1} shows the JESs, NESs and PESs for dissociative ionization of \mol for laser pulses with $\tau=1.1$ fs 
and two different intensities. 
In the JES panels, we define the Stark energy shift $\Delta$ of the ground state $\ket{\Psi_0}$ to be the shift at the field maximum, $t=0$, [see Eqs.~\eqref{eq:vecpot} and \eqref{eq:las_env}]
\begin{equation}
  \begin{aligned}
    \Delta=\dot{\phi}(0)-E_0,
    \label{eq:tdse_stark}
  \end{aligned}
\end{equation}
where $\phi(t)$ is the phase of the ground-state amplitude $\braket{\Psi_0}{\Psi(t)}=\abs{\braket{\Psi_0}{\Psi(t)}}e^{-i\phi(t)}$, which is directly extracted from the TDSE calculations \cite{Demekhin13}.
In the side insets of Fig.~\ref{fig1}, the NESs from the TDSE calculations and the reflection principle are shown. The reflection principle \cite{Chelkowski99,Barmaki07,Schmidt12} amounts to the approximation where the electron 
is emitted into the continuum by the laser at the internuclear distance $R$, leaving behind two bare protons that Coulomb explode, gaining the kinetic energy $\en=1/R$. The NES is then obtained by reflecting the probability density of the initial vibrational state $\chi_0(R)$, and weighting with $-dR/d\en=1/E_\text{N}^2$:
\begin{equation}
  \frac{dP}{dE_\text{N}}\propto \frac{\abs{\chi_0(1/E_\text{N})}^2}{E_\text{N}^2}.
  \label{eq:tdse_reflection}
\end{equation}
In the upper insets of Fig.~\ref{fig1}, the PESs are shown for the moving-nuclei and fixed-nuclei TDSE calculations. For the fixed-nuclei calculations, we fix the internuclear distance at the equilibrium $R_0 \equiv \bra{\Psi_0}R\ket{\Psi_0}=2.06$.

For the intensity $I=3\times 10^{17}$ W/cm$^2$ in Fig.~\ref{fig1}(a), the JES displays a single peak centered at $\left(E_\text{e},E_\text{N}\right)=(1.23,0.46)$, with a region of zero density along the line $E_\text{N}=0.6$ (perhaps more discernible in the NES). In the NES panel, the result for the reflection approximation peaks at $E_\text{N}=1/R_0=0.484$, a clear shift with respect to the correct NES peak located at 0.455. Furthermore, the minimum at $E_\text{N}=0.6$ is absent in the reflection principle result. This is to be expected, as the reflection principle is a crude approximation, and can be used only for qualitative analysis. In the PES panel, there is a shift of the peak of the fixed-nuclei result compared to the moving-nuclei result. This shift is due to the fact that the fixed-nuclei result does not take into account the probability density of the initial vibrational state $\abs{\chi_0(1/E_\text{N})}^2$. 

For the more intense pulse in Fig.~\ref{fig1}(b), additional structures appear in the JES and the Stark shift is larger as indicated by the dashed energy conservation lines in the JESs. At least four peaks are now clearly visible in the JESs, with the largest one centered at $\left(E_\text{e},E_\text{N}\right)=(1.28,0.44)$. The three most visible peaks are more or less along the energy conservation lines $E_\text{e}+E_\text{N}=1.62,1.72,1.85$ [lines not drawn in Fig.~\ref{fig1}(b)].
In the NESs, the TDSE result is shifted towards lower $\en$, with the magnitude of the shift similar to that in the case for the lower laser intensity in Fig.~\ref{fig1}(a).
In the PESs for moving nuclei, only two peaks are visible, with the highest-energy peak in the JESs being weighted out from the integration of the JESs. We therefore stress the importance of the JESs: if only the PESs and the NESs were at our disposal, no information on the third peak in the JESs could be obtained. This extends the conclusion of Refs.~\cite{Madsen12,Silva13,Wu13} that the JES is a very useful observable to the XUV regime.

It should be noted that in Ref.~\cite{Demekhin13b}, it was 
suggested
that the combination of using small simulation volumes and CAPs placed at the box boundaries would make it impossible to produce dynamic interference in the PES, and enormously large simulation volumes (radial coordinate up to  $r_\text{max}= 10\:000$) were used in their calculations to obtain the PESs for hydrogen. However, as shown in Fig.~\ref{fig1}(b),
it is indeed possible to observe interference effects in the JESs and PESs by using t-SURFF in a small simulation volume $\abs{x}\le 200$.

\begin{figure}
  \centering
  \includegraphics[width=0.45\textwidth]{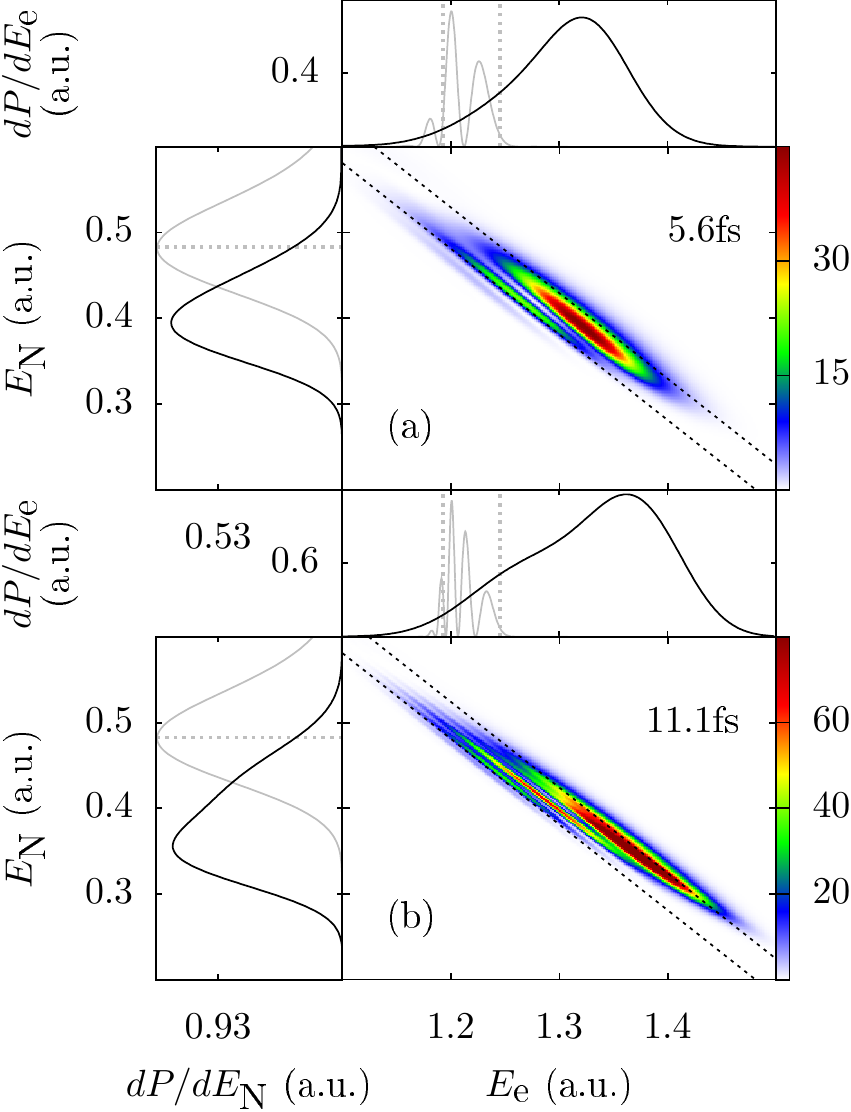}
  \caption{(Color online)
    As Fig.~\ref{fig1}, but for pulses with parameters $\omega=2.278$, $I=15\times 10^{17}$ W/cm$^2$, and two different pulse durations.}
  \label{fig2}
\end{figure}

Figure~\ref{fig2} shows the JESs for longer laser pulses $\tau=5.6$ fs and $11.1$ fs, with intensity $I=15\times 10^{17}$ W/cm$^2$. 
The increase of the pulse duration has several effects on the JESs. First, due to the smaller bandwidth of the laser pulse, the JESs in Fig.~\ref{fig2} are now narrower
compared to the results for the shorter pulses in Fig.~\ref{fig1}. Second, the increase of pulse duration from 5.6 fs in Fig.~\ref{fig2}(a) to 11.1 fs in Fig.~\ref{fig2}(b) leads to more interference peaks emerging, similar to the case of dynamic interference in hydrogen \cite{Demekhin12,Demekhin13,Demekhin13b,Yu13a}, but now visible along the diagonal in the JES. 
In addition, the JES and NES for the $5.6$ fs pulse in Fig.~\ref{fig2}(a) are shifted toward smaller $E_\text{N}$ values compared to the shorter pulses used in Fig.~\ref{fig1}, with the peak now located around $E_\text{N}=0.4$ in the NES.
For the $11.1$ fs pulse in Fig.~\ref{fig2}(b), a peak is observed around $E_\text{N}=0.36$ in the NES, while a ``shoulder'' structure is seen around $E_\text{N}=0.42$. 
When we compare the PES results for moving nuclei to the corresponding fixed-nuclei results in Fig.~\ref{fig2}, the PES for moving nuclei has its dynamic interference peaks completely smeared out, with no interference patterns visible. Furthermore, the PES for moving nuclei in Fig.~\ref{fig2}(b) peaks at $\ee=1.31$, a clear shift with respect to the fixed-nuclei peak around $\ee=1.21$. Thus, due to the inclusion of the nuclear degree of freedom, the fixed-nuclei results for the PESs are completely wrong. This again stresses the importance of using the JESs for the detection of dynamic interference in molecules.
  
\section{Analysis of the JES}
\label{sec:analysis}
We will now analyze the structures in the spectra of 
Figs.~\ref{fig1} and \ref{fig2}, using two methods.
\subsection{Essential-states expansion}
\label{sec:analysis1}
In the first analysis we follow Ref.~\cite{Demekhin12} and extend it to the molecular case of \mol by including the nuclear degree of freedom.
The molecule-laser interaction is now chosen in the length gauge, where the interaction potential in Eq.~\eqref{eq:Hamiltonian} is given by $V_\text{I}^\text{LG}=\beta_\text{LG}xF(t)$, with $\beta_\text{LG}=1+1/(2m_\text{p}+1)$. The electric field is chosen in the form $F(t)=F_0 g(t)\cos(\omega t)$, with $g(t)$ given by Eq.~\eqref{eq:las_env} and $F_0=\omega A_0$. For the many-cycle pulses considered in this work, the carrier-envelope phase difference between the fields in the length and velocity gauges is unimportant for the resulting spectrum \cite{Madsen02}. We also verified this in fixed-nuclei TDSE calculations where we checked that the usage of length and velocity gauge Hamiltonians produces identical spectra.

The wave function is first expanded in terms of the ``essential'' states consisting of the initial state and the continuum eigenstates of the field-free Hamiltonian
\begin{equation}
  \begin{aligned}
  \ket{\Psi(t)}=&c_0(t)\ket{\Psi_0}e^{-iE_0t}\\
  &+\sum_P\int d{E_\text{e}}\int d{E_\text{N}}c^{P}_{E_\text{e},E_\text{N}}(t)\ket{u^{P}_{E_\text{e},E_\text{N}}}e^{-i\omega t},
  \label{eq:essential_expansion}
  \end{aligned}
\end{equation}
where $\ket{u^P_{E_\text{e},E_\text{N}}}$ is a field-free continuum state of \mol with parity $P$ (``e'' for even, ``o'' for odd), electronic continuum energy $E_\text{e}$ and nuclear continuum energy $E_\text{N}$. The latter states are obtained in the BO approximation by the method outlined in the Appendix \ref{appendix1}.
Inserting Eq.~\eqref{eq:essential_expansion} into the TDSE \eqref{eq:TDSE} and projecting onto the ``essential'' states, we arrive at the coupled differential equations:
\begin{subequations}
  \begin{equation}
%  \begin{eqnarray}
    i\dot{c_0}(t)=\int d{E_\text{e}}\int d{E_\text{N}} \left[\frac{1}{2}{d_{E_\text{e},E_\text{N}}^{\text{o}*}}F_0\right]g(t)e^{iE_0t}c_{E_\text{e},E_\text{N}}^{o}(t) \label{eq:sec4_2a}
  \end{equation}
  \begin{equation}
    \begin{aligned}
      i\dot{c}_{E_\text{e},E_\text{N}}^{o}(t)=&\left[\frac{1}{2}{d_{E_\text{e},E_\text{N}}^{\text{o}}}F_0\right]g(t)e^{-iE_0t}c_0(t)\\
    &+\left(E_\text{e}+E_\text{N}-\omega\right)c^\text{o}_{E_\text{e},E_\text{N}}(t),\label{eq:sec4_2b}
  \end{aligned}
  \end{equation}
%  \end{eqnarray}
  \label{eq:sec4_2}
\end{subequations}
where $d_{E_\text{e},E_\text{N}}^{P}=\bra{u^{P}_{E_\text{e},E_\text{N}}}x\ket{\Psi_0}$ is the transition dipole matrix element. In obtaining Eq.~\eqref{eq:sec4_2}, we have used the RWA and the fact that the initial state $\ket{\Psi_{0}}$ has even parity. Furthermore, we have neglected the continuum-continuum couplings.
The solution to 
Eq.~\eqref{eq:sec4_2b} is given by
\begin{equation}
  \begin{aligned}
  c_{\ee,\en}^{\text{o}}(t)=&-i\left[\frac{1}{2}d_{\ee,\en}^\text{o} F_0\right]
  e^{-i (\delta+E_0) t}\\
  &\times \int_{-\infty}^tc_0(t')g(t')e^{i\delta t'}dt',
  \label{eq:essential_continuum}
  \end{aligned}
\end{equation}
where
\begin{equation}
  \label{eq:essential_detuning}
  \delta=\ee+\en-\omega-E_0
\end{equation}
is the detuning.
The JES is then
\begin{equation}
  \frac{\partial^2P}{\partial E_\text{e}\partial E_\text{N}}=\abs{c_{E_\text{e},E_\text{N}}^{\text{o}}(T)}^2
  \label{eq:essential_jes}
\end{equation}
with $T\gg \tau$.

\begin{figure}
  \centering
  \includegraphics[width=0.45\textwidth]{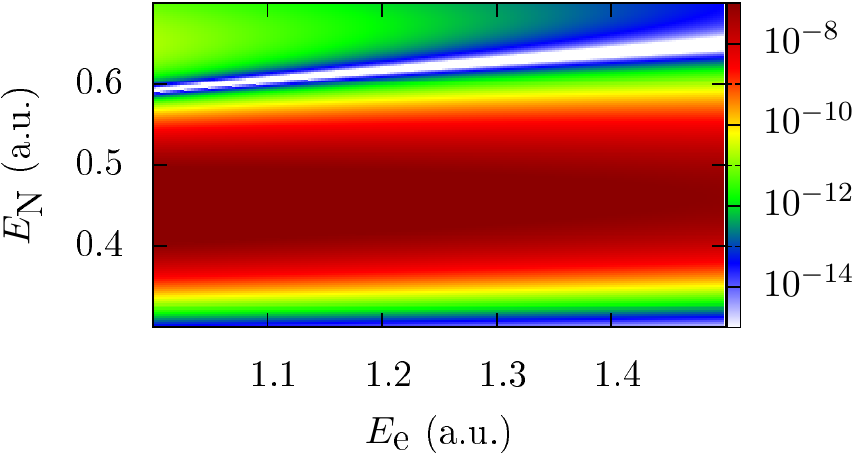}
  \caption{(Color online) The transition dipole matrix element squared  $\abs{d_{E_\text{e},E_\text{N}}^{\text{o}}}^2=\abs{\bra{u^{\text{o}}_{E_\text{e},E_\text{N}}}x\ket{\Psi_0}}^2$ [see the discussion of Eq.~\eqref{eq:sec4_2}].}
  \label{fig3}
\end{figure}

We see from Eqs.~\eqref{eq:essential_continuum} and \eqref{eq:essential_jes} that in the present model, the structure of $\abs{d_{E_\text{e},E_\text{N}}^{\text{o}}}^2$ will determine the structure of the JES. The former is plotted in Fig.~\ref{fig3}. The highest-density region is located along $E_\text{N}=0.45$, immediately explaining the peaks in the JESs and NESs of Fig.~\ref{fig1}(a). Furthermore, a valley of zero density is seen in Fig.~\ref{fig3}, explaining the minima in Fig.~\ref{fig1}(a) located around $\left(E_\text{e},E_\text{N}\right)=(1.1,0.6)$ in the JESs. However, Fig.~\ref{fig3} does not explain the structures in Fig.~\ref{fig1}(b), where the peak in the NES is still at $E_\text{N}=0.45$, but instead of a minimum at $E_\text{N}=0.60$, there is now a local maximum.

To calculate the JESs using Eq.~\eqref{eq:essential_jes}, we need to solve the coupled differential equations in Eq.~\eqref{eq:sec4_2}. These equations can be decoupled by following the procedure of Refs.~\cite{Demekhin12,Demekhin13}. First, if $c_0(t')g(t')$ in Eq.~\eqref{eq:essential_continuum} varies slowly compared to the rest of the integrand, it can be taken out of the integral and evaluated at time $t$. This approximation will be referred to as the local approximation. 
  By evaluating Eq.~\eqref{eq:essential_continuum}, plugging the result into Eq.~\eqref{eq:sec4_2a}, and solving the resulting uncoupled differential equation, we obtain $c_0(t)$ in the form 
\begin{equation}
  c_0(t) \cong e^{-\left(i\Delta+\Gamma/2\right)J(t)},
  \label{eq:essential_ground}
\end{equation}
where $\Delta$ is the Stark shift, $\Gamma$ the ionization rate, and 
\begin{equation}
  \label{eq:essential_J}
  J(t)=\int_{-\infty}^{t}g^2(t')dt'.
\end{equation}
As noted in Refs.~\cite{Demekhin12,Demekhin13}, $\Delta$ is difficult to calculate and also depends on the ``non-essential'' states which were omitted in the present model. 
We therefore follow the procedure proposed in Ref.~\cite{Demekhin13} and extract $\Delta$ from the TDSE calculations by employing Eq.~\eqref{eq:tdse_stark} and use it in Eq.~\eqref{eq:essential_ground}. The Stark shift obtained in this way is beyond the RWA. 
The expression for $\Gamma$ obtained in the derivation of Eq.~\eqref{eq:essential_ground} is
\begin{equation}
  \Gamma=2\pi \int d\ee\int d\en \abs{\frac{d_{\ee,\en}F_0}{2}}^2 \delta\left( E_0+\omega-\ee-\en \right).
  \label{eq:demekhin_gamma}
\end{equation}
Note that Eq.~\eqref{eq:demekhin_gamma} can also be obtained using Fermi's golden rule.
Compared with the previous works on atoms \cite{Demekhin12,Demekhin13}, $\Gamma$ now contains an extra integral over the nuclear kinetic energy, with the energies related by $\ee+\en=E_0+\omega$.
For the laser pulse used in Fig.~\ref{fig1}(a), the survival probability of the ground state in the present model is
\begin{equation}
  P_0^\text{model}(\infty)= \abs{c_0(\infty)}^2=e^{ -\Gamma J(\infty)}=0.978,
\end{equation}
which is comparable to the TDSE result $P_0^\text{TDSE}=0.975$. This demonstrates that the approximation \eqref{eq:demekhin_gamma} is not too bad, at least for the intensity used in Fig.~\ref{fig1}(a).

\begin{figure}
  \centering
  \includegraphics[width=0.45\textwidth]{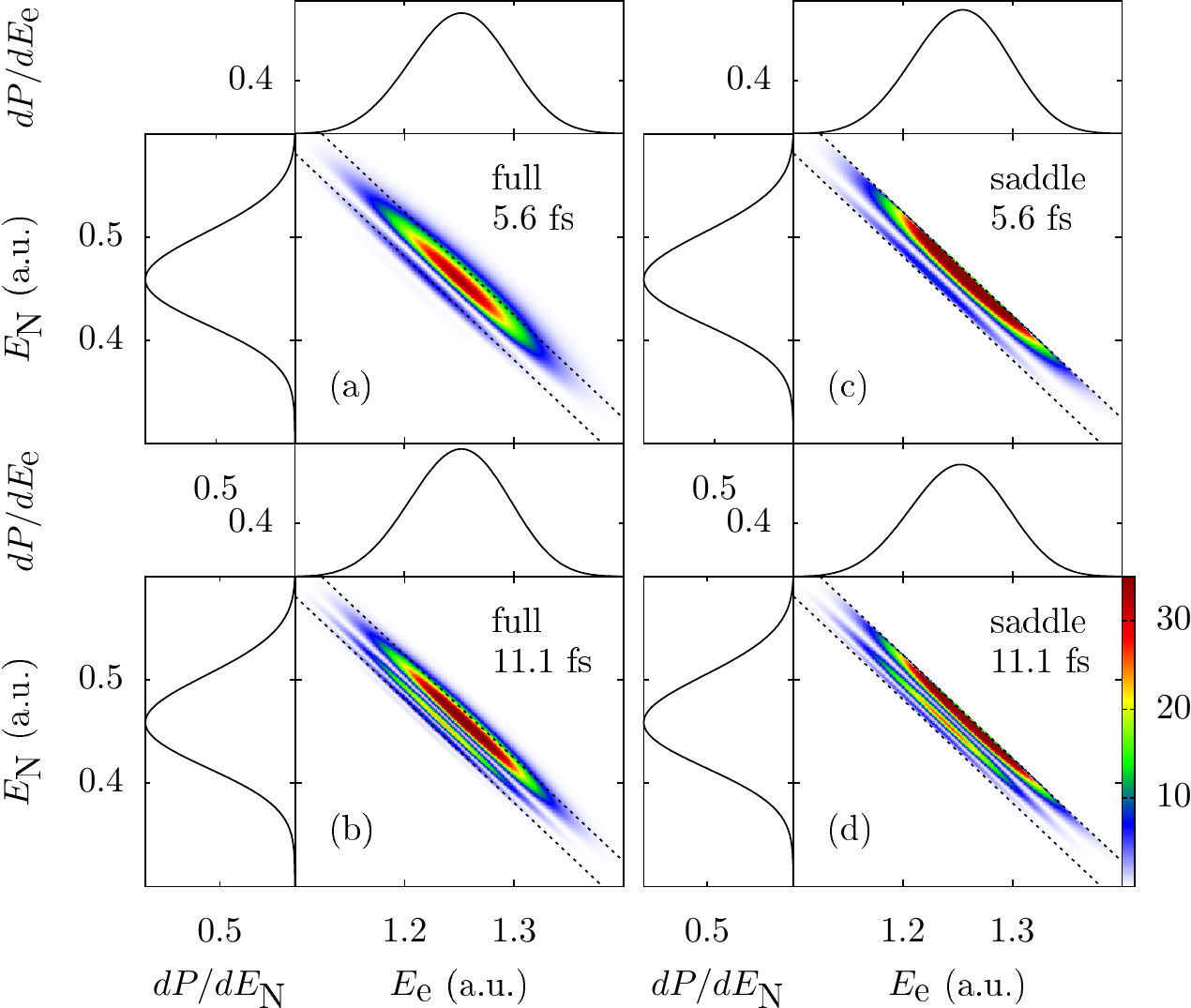}
  \caption{(Color online) (a),(b): JESs calculated using the approximation \eqref{eq:essential_jes_approx} (denoted by ``full'' in the panels). (c),(d) JESs calculated using the saddle-point approximation \eqref{eq:essential_jes_saddle} (denoted by ``saddle'' in the panels). Laser parameters are the same as in Fig.~\ref{fig2}, i.e., $\omega=2.278$, $I=24\times 10^{17}$ W/cm$^2$, and two different pulse durations $\tau$ displayed in the figure. The magnitudes of the spectra are in arbitrary units. The left and right dashed diagonal lines correspond to $\delta=0$ and $\delta=\Delta$, respectively.}
  \label{fig4}
\end{figure}

We may insert the approximation for $c_0(t)$ given by Eq.~\eqref{eq:essential_ground} into Eq.~\eqref{eq:essential_continuum} to obtain an approximate formula for the JES in the local approximation:
\begin{equation}
  \frac{\partial^2P}{\partial E_\text{e}\partial E_\text{N}} \cong
  \abs{\frac{1}{2}d_{\ee,\en}^\text{o} F_0 \int_{-\infty}^Tg(t')e^{-\Gamma/2 J(t')+i\Phi(t')}dt'}^2,
  \label{eq:essential_jes_approx}
\end{equation}
with
\begin{equation}
  \Phi(t)=\delta t-\Delta J(t).
  \label{eq:essential_phase}
\end{equation}
It should be noted that because $c_0(t)$ of Eq.~\eqref{eq:essential_ground} was obtained in the local approximation, Eq.~\eqref{eq:essential_jes_approx} also works only in this approximation, a point we will come back to below.
Figures~\ref{fig4}(a) and \ref{fig4}(b) show the JESs, calculated using Eq.~\eqref{eq:essential_jes_approx}, and their corresponding PESs and NESs, for two different pulse durations (the same laser parameters as in Fig.~\ref{fig2}). In the JES of Fig.~\ref{fig4}(a), two peaks are clearly visible, with a large peak located along $\ee+\en=1.71$ and a smaller one along $\ee+\en=1.68$. This is in agreement with the results for the full TDSE calculation shown in Fig.~\ref{fig2}(a), where we also had two clearly visible peaks along the same mentioned diagonals. However, in the present model where the ground state is assumed to be the only bound state, the norm squared of the dipole matrix element  $\abs{d_{\ee,\en}^{\text{o}}}^2$ 
%in the sense that the number of peaks, the relative amplitude of the peaks, and the positions of the peaks are along  
restricts the NES to be centered around $\en=4.5$, in disagreement with Fig.~\ref{fig2}(a).
In the JES of Fig.~\ref{fig4}(b), three peaks are now visible, along the diagonal lines $\ee+\en=1.68,1.69,1.71$. 
The number of peaks and diagonal positions of these peaks are in agreement with the full TDSE result in Fig.~\ref{fig2}(b). The spectrum is again incorrectly centered around $\en=4.5$, because of $\abs{d_{\ee,\en}^{\text{o}}}^2$ and the insufficient description of the nuclear degree of freedom.

To gain more physical insight into the structures of the JESs in Figs.~\ref{fig4}(a) and \ref{fig4}(b), we have evaluated Eq.~\eqref{eq:essential_jes_approx} using the saddle-point approximation and in this way obtained an approximate expression for the JESs (see Refs.~\cite{Demekhin12,Demekhin13} for a similar expression for the PESs in the atomic case):
\begin{equation}
  \begin{aligned}
    \frac{\partial^2P}{\partial \ee\partial \en}
    \propto&
    \abs{ \frac{d_{\ee,\en}^\text{o} F_0}{2} \sum_{t=\pm t_s} \frac{g(t)}{\sqrt{\abs{\ddot{\Phi}(t)}}} e^{-\Gamma/2 J(t)}e^{i\left[\Phi(t)+\zeta(t)\right]}}^2\\
    \propto&
    \abs{ \frac{d_{\ee,\en}^\text{o} F_0 \tau }{2\sqrt{\Delta t_s}} }^2
    \left\{
      e^{-\Gamma J(-t_s)}+e^{-\Gamma J(t_s)}\right.\\
    &+\left.2e^{-\Gamma/2 \left[J(-t_s)+J(t_s)\right]}
%      \cos\left[-2\delta t_s-\Delta (J(-t_s)-J(t_s))-\pi/2\right]
      \cos\left[ K(\delta) \right]
    \right\},
    \label{eq:essential_jes_saddle}
  \end{aligned}
\end{equation}
with
\begin{equation}
  \label{eq:essential_jes_K}
  K(\delta)=-2\delta t_s-\Delta (J(-t_s)-J(t_s))-\pi/2.
\end{equation}
In Eqs.~\eqref{eq:essential_jes_saddle} and \eqref{eq:essential_jes_K}, $ \zeta(\pm t_s)=\pm \pi/4$; $J(t)$ is given by Eq.~\eqref{eq:essential_J}; and $\pm t_s$ are the saddle points with
\begin{equation}
  \begin{aligned}
    t_s=\frac{\tau}{2}\sqrt{\frac{\ln(\Delta/\delta)}{2\ln(2)}}
  \end{aligned}
\end{equation}
satisfying the stationary-phase condition
\begin{equation}
  \dot{\Phi}(\pm t_s)=\delta-\Delta g^2(\pm t_s)=0.
  \label{eq:essential_saddle_condition}
\end{equation}
The JESs calculated using the saddle-point expression \eqref{eq:essential_jes_saddle} are shown in Figs.~\ref{fig4}(c) and \ref{fig4}(d).
Comparing these results with Figs.~\ref{fig4}(a) and \ref{fig4}(b), it is seen that the saddle-point approximation captures the essential features of the JESs calculated using the full time-integral \eqref{eq:essential_jes_approx}: the number of peaks perpendicular to the diagonal, the positions of the peaks, and the relative intensities of the peaks are qualitatively similar to the results of the full calculation. 
From the stationary-phase condition \eqref{eq:essential_saddle_condition} and the definition of $g(t)$ in Eq.~\eqref{eq:las_env}, it is seen that for a saddle point to exist, the detuning must belong to the interval $\delta\in \left[0,\Delta\right]$. For $\delta$ lying in this interval, $K(\delta)$ is a monotonically decreasing function, as $dK/d\delta=-2\delta \le 0$. Furthermore, $K(0)=\Delta J(\infty)-\pi/2$ and $K(\delta)=-\pi/2$.  %From Eq.~\eqref{eq:essential_jes_K} and the expression for $J(t)$ in Eq.~\eqref{eq:essential_J}, $K(\delta)=\sqrt{\pi/(8\ln2)\Delta \tau-\pi/2}$ 
The total accumulated phase of the cosine in Eq.~\eqref{eq:essential_jes_saddle} is thus $\Delta J(\infty)=\sqrt{\pi/(8\ln2)}\Delta\tau \equiv 2\pi k$, with $k$ corresponding to the number of oscillations. 
%As $\cos[K(\delta)]=0$, $k$ rounded down to the nearest integer will correspond to the number of minimas 
For $\tau=5.6$ fs the accumulated phase is $1.19\times 2\pi$ while for $\tau=11.1$ fs it is $2.38 \times 2\pi$. These numbers are indeed in qualitative agreement with the JESs of Figs.~\ref{fig4}(c) and ~\ref{fig4}(d), where slightly more than one oscillation and slightly more than two oscillations are visible, respectively.
At $\Delta=\delta$, the saddle point is $t_s=0$, and the approximation for the JES in Eq.~\eqref{eq:essential_jes_saddle} diverges.
%This divergence manifests in Fig.~\ref{fig4} near the dashed lines corresponding to $\Delta=\delta$ as the discrepancy between the full results in Figs.~\ref{fig4}(a) and \ref{fig4}(b), and the saddle-point results in Figs.~\ref{fig4}(c) and \ref{fig4}(d), .
This divergence is seen in Fig.~\ref{fig4} as the discrepancy between the ``full'' results and the ``saddle'' results near the right dashed lines corresponding to $\Delta=\delta$.
We have thus demonstrated that the interference patterns in the JESs are indeed due to the dynamic Stark effect, where two contributions to the same energy pair $(\ee,\en)$ in the JES coherently superimpose. Figure~\ref{fig4} again confirmed that the introduction of a nuclear degree of freedom makes the dynamic interference effect completely invisible in the PESs.
% For the intense short pulse with $\color{blue}\tau=1.1$ fs and $24\times 10^{17}$ W/cm$^2$ considered in Sec.~\ref{sec:num_res} the local approximation used to decouple the Eqs.~\eqref{eq:sec4_2} to obtain the approximate expression for $c_0$ \eqref{eq:essential_ground} will fail, and we choose not to make the analysis in this section for that pulse.}

\begin{figure}
  \centering
  \includegraphics[width=0.4\textwidth]{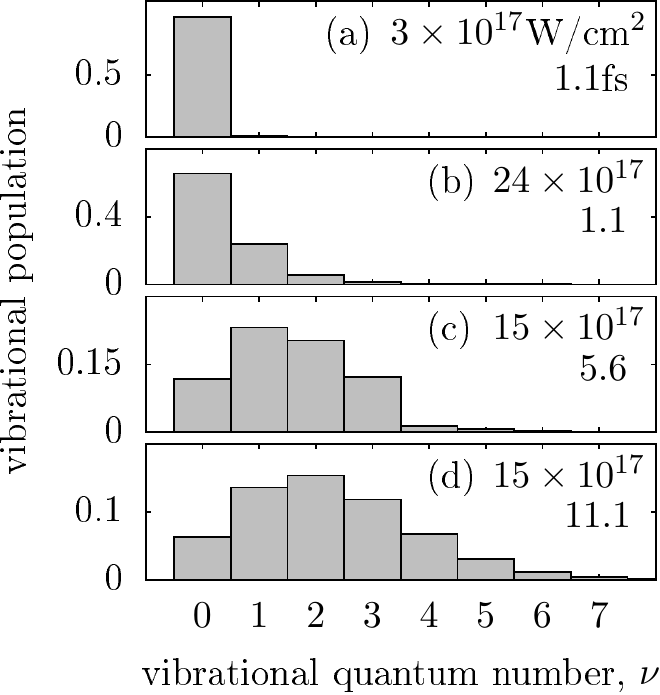}
  \caption{(Color online) Vibrational distributions at the end of the laser pulses, for the laser parameters used in Figs.~\ref{fig1} and \ref{fig2}.}
  \label{fig5}
\end{figure}

We now discuss the two main deficiencies in the present model.
The first one is the neglect of the excited vibrational states in the expansion of Eq.~\eqref{eq:essential_expansion}, which could have been populated during the pulse by impulsive Raman-type transitions from the ground vibrational state. 
The vibrational populations of the first few vibrational states at the end of the laser pulse are shown in Fig.~\ref{fig5}. For the short pulses considered in Figs.~\ref{fig5}(a) and \ref{fig5}(b), the population is predominantly in the vibrational ground state, and the expansion in Eq.~\eqref{eq:essential_expansion} is expected to be a good approximation. However, for the high-intensity pulse with $I=24\times 10^{17}$ W/cm$^2$, the variation of $c_0(t')g(t')$ in Eq.~\eqref{eq:essential_continuum} is comparable to the rest of the integrand, and the local approximation cannot be applied. As a result, the approximation for the JESs in Eqs.~\eqref{eq:essential_jes_approx} and \eqref{eq:essential_jes_saddle} will fail. 
%which is the reason why we chose not to make the analysis in this section for this particular pulse. 
For the longer pulses, the local approximation is expected to work well. However, the pulse energy of the laser is much greater, meaning that much stronger Raman couplings and thus much greater population of excited vibrational states is observed, as shown in Figs.~\ref{fig5}(c) and \ref{fig5}(d). The essential expansion in Eq.~\eqref{eq:essential_expansion} that neglects the excited vibrational states is thus expected to break down, which is indeed seen to be the case by comparing the results of Figs.~\ref{fig2} and \ref{fig4}. Although Eq.~\eqref{eq:essential_expansion} is not expected to work well, the analysis in the present section is still useful, as it provides insight into the structures of the JES in terms of dynamical interference, provided that only the vibrational ground state is populated.
%This is easily seen by comparing the transition dipole matrix element in Fig.~\ref{fig3} with the JES in Fig.~\ref{fig2}, e.g., by noticing that there is no peak at $\en=0.45$ in Fig.~\ref{fig2}.

Another deficiency in the present model is the neglect of continuum-continuum couplings.
The ponderomotive energy of the most intense pulse used is $U_\text{p}=3.29$, larger than the photon energy $\omega=2.278$. This indicates that multiphoton processes at these intensities cannot be neglected and will contribute to differences in the JESs calculated using the present model and the JESs obtained from the full TDSE calculations. Such differences were observed in the atomic case \cite{Demekhin13b} in terms of differences in the relative amplitudes of the peaks in the PES.

\subsection{Reflection method for the JES}
In situations where ionization from the initial state is dominant, the JES can be approximated by weighting the fixed-nuclei results with the initial vibrational density:
\begin{equation}
  \begin{aligned}
    \pdt{P}{E_\text{e}}{E_\text{N}} \propto \frac{\abs{\chi_0\left(1/E_\text{N}\right)}^2}{E_\text{N}^2}\frac{dP}{dE_\text{e}}\left(1/E_\text{N}\right),
    \label{eq:jesweight}
  \end{aligned}
\end{equation}
where $dP/dE_\text{e}(R)$ is the PES calculated at the fixed internuclear distance $R$. 
We refer to this model as the reflection method for the JES.
\begin{figure}
  \centering
  \includegraphics[width=0.5\textwidth]{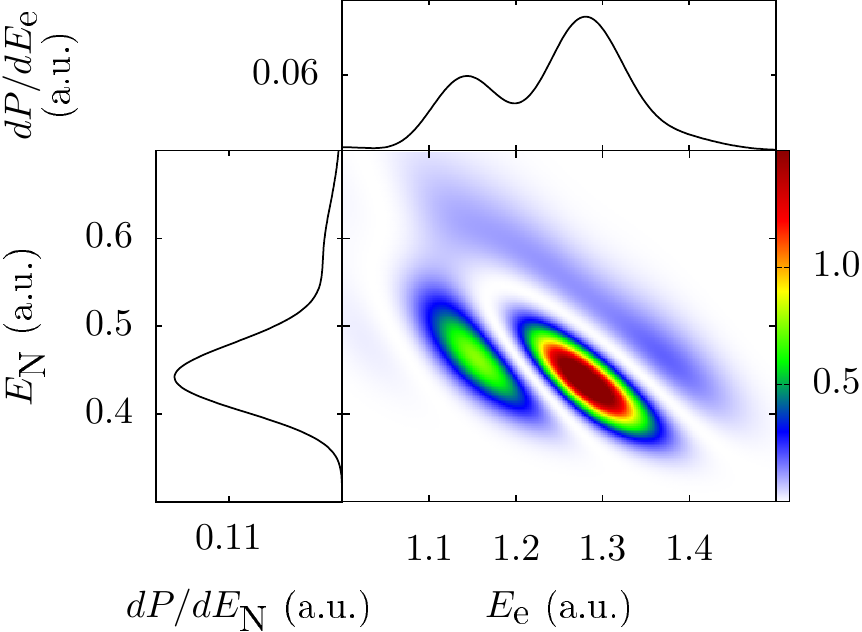}
  \caption{(Color online) JES calculated using Eq.~\eqref{eq:jesweight} for the same laser parameters as in Fig.~\ref{fig1}(b), $\omega=2.278$, $\tau=1.1$ fs, and $I=24\times 10^{17}$ W/cm$^2$.
}
  \label{fig6}
\end{figure}
\begin{figure}
  \centering
  \includegraphics[width=0.5\textwidth]{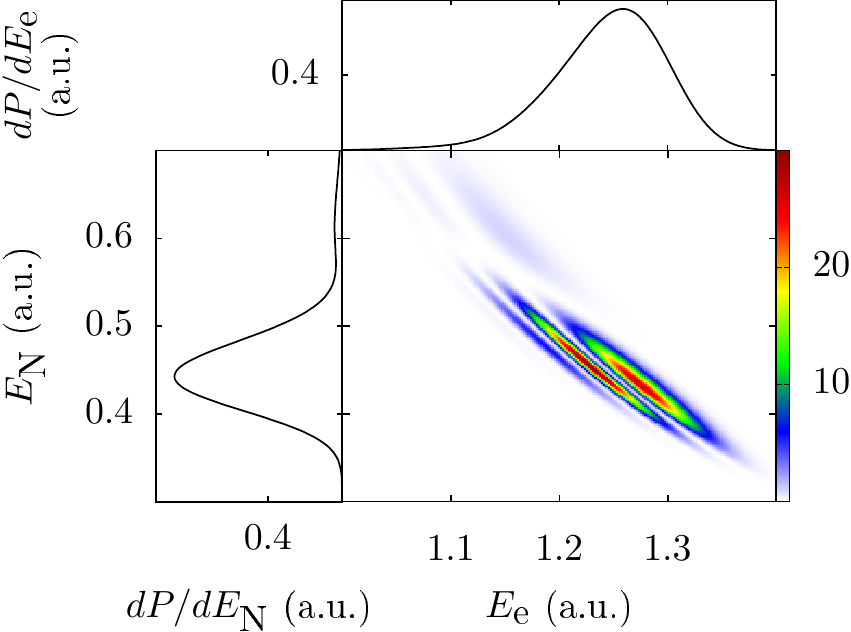}
  \caption{(Color online) JES calculated using Eq.~\eqref{eq:jesweight} for the same laser parameters as in Fig.~\ref{fig1}(a), $\omega=2.278$, $\tau=5.6$ fs, and $I=15\times 10^{17}$ W/cm$^2$.}
  \label{fig7}
\end{figure}
The JESs calculated using Eq.~\eqref{eq:jesweight} for the field parameters in Fig.~\ref{fig1}(b) are shown in Fig.~\ref{fig6}. There is indeed a good qualitative match, with all structures in the JESs of Fig.~\ref{fig1}(b) accounted for in Fig.~\ref{fig6}. The reason for this is simple: all of the electron-laser couplings are included in $dP/dE_\text{e}(R)$ of Eq.~\eqref{eq:jesweight}, while there are minimal populations of higher excited vibrational states (see Fig.~\ref{fig5}).

For the longer pulses considered in Fig.~\ref{fig2} there is significant vibrational excitation during the pulse, as shown in 
Figs.~\ref{fig5}(c) and \ref{fig5}(d), and
 the approximation leading to Eq.~\eqref{eq:jesweight} is not valid. This is indeed verified by comparing Fig.~\ref{fig7} with Fig.~\ref{fig2}(a), where in Fig.~\ref{fig2}(a) the JES is shifted towards lower $\en$.
Notice that we have solved the TDSE exactly at fixed internuclear distances in the present approximation, meaning that the electronic continuum-continuum couplings neglected in the ``essential states'' model from Sec.~\ref{sec:analysis1} are included here.

\begin{figure}
  \centering
  \includegraphics[width=0.4\textwidth]{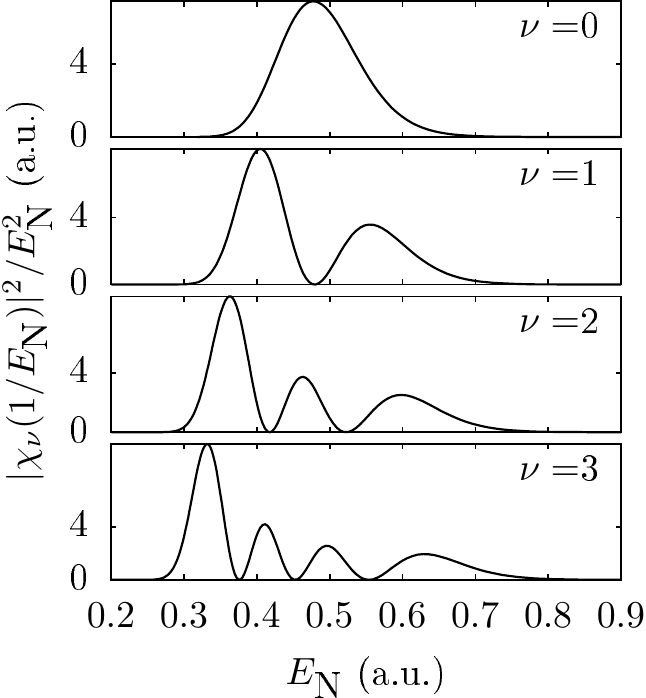}
  \caption{Reflection results [Eq.~\eqref{eq:tdse_reflection}] for the lowest four vibrational states.}
  \label{fig8}
\end{figure}

The shift towards lower nuclear energies in the JESs of Fig.~\ref{fig2} compared to Fig.~\ref{fig1}
is due to the excited vibrational populations, which can be shown qualitatively using the simple reflection principle \eqref{eq:tdse_reflection}. In Fig.~\ref{fig8}, results of Eq.~\eqref{eq:tdse_reflection} for the lowest four vibrational states are plotted.
For the $\tau=5.6$ fs pulse in Fig.~\ref{fig2}(a), the most populated vibrational state at the end of the pulse is $\nu=1$. In Fig.~\ref{fig8}, the reflection result for $\nu=1$ has the large peak at $\en=0.408$, thus explaining the peak in the NES of Fig.~\ref{fig2}(a) at around $\en=0.408$.
Similarly, for the $\tau=11.1$ fs pulse in Fig.~\ref{fig2}(b), the most populated vibrational state is $\nu=2$, which in Fig.~\ref{fig8} has the largest peak located at $\en=0.366$, explaining the peak in the NES of Fig.~\ref{fig2}(b) at around $E_\text{N}=0.36$. The shoulder structure in the NES of Fig.~\ref{fig2}(b) at around $E_\text{N}=0.42$ can be interpreted as resulting from the $\nu=1$ state.

\section{Conclusion}
\label{sec:conclusion}
We investigated dissociative ionization of \mol using intense, femtosecond XUV laser pulses by propagating the TDSE for a collinear model of \mol. 
The molecular t-SURFF method \cite{Yue13} was employed to obtain the JESs and was shown to work well in the present high-frequency, high-intensity regime. Using this method, we were able to significantly reduce the simulation volume and thereby the computational effort. The dynamic interference effect, which was first observed in theoretical calculations for simple atomic systems \cite{Toyota07,Demekhin12}, was shown to be present in the case of \mol as well, emerging as interference structures in the JESs. The PESs and NESs were shown to be inadequate for the observation of the dynamic interference effect in \mol. For the longer pulses (11.1 fs) used, a clear shift of the NESs and JESs toward lower nuclear kinetic energies was observed.

We analyzed the resulting JESs in terms of two different models. In the first model, we expanded the wave function in terms of the ``essential'' states of the system, consisting of the ground state and the continuum states. In the RWA and local approximation, and neglecting the continuum-continuum couplings, the formulation for the JES was expressed in terms of the Stark shift and the ionization rate.
The Stark shift was extracted from TDSE calculations to take into account effects of the nonessential states. 
The model was shown to produce interference structures in the JESs, and by making a saddle-point approximation, we were able to relate the structures to the dynamic interference effect. However, due to the neglect of the excited vibrational states in the model, the JESs calculated for the longer laser pulses were shifted towards smaller values of $\en$ and larger values of $\ee$ compared to the full TDSE results.

In the second model, we calculated the PESs of \mol at fixed internuclear distances and obtained the JESs from simple reflection arguments.
Although this model was able to produce the correct JESs for the short pulses used, it was unable to produce the correct JESs for the longer pulses, due to the involvement of higher vibrational states excited during the pulse. However, by using the reflection principle on the excited nuclear vibrational states, we could qualitatively explain the shift of the JESs towards lower nuclear kinetic energies.

\begin{acknowledgments}
We thank J. Svensmark and J. E. B{\ae}kh{\o}j for a careful reading of the manuscript, and C. Yu 
for useful discussions.
This work was supported by the Danish Center for Scientific Computing, an ERC-StG (Project No. 277767---TDMET), and the VKR Center of Excellence, QUSCOPE.
\end{acknowledgments}

\appendix*
\section{Field-free continuum eigenstates}
\label{appendix1}
In this appendix, we describe 
the method used for obtaining the field-free continuum states of \mol in the BO approximation. 

The time-independent Schr\"{o}dinger equation reads
\begin{equation}
  \label{eq:app1_tise}
  \bigl[T_\text{e}+T_\text{N}+V_\text{eN}(x,R)+V_\text{N}(R)\bigr]u(x,R)=E u(x,R).
\end{equation}
The continuum states $u(x,R)$ of Eq.~\eqref{eq:app1_tise} depend on the electronic energies $E_\text{e}$, the nuclear energies $E_\text{N}$, and the parity $P$. We make in Eq.~\eqref{eq:app1_tise} the ansatz
\begin{equation}
  \floqu{E_\text{e}E_\text{N}}{P}=\floqcel{x;R}\floqcnuc{R},
\end{equation}
and by neglecting the action of $T_\text{N}$ on $\floqcel{x,R}$ (BO approximation), the electronic and nuclear degrees of freedom decouple, resulting in the equations
\begin{align}
  \bigl[ T_\text{e}+V_\text{eN}(x,R) \bigr]\floqcel{x;R}&=E_\text{e}\floqcel{x;R},\label{eq:HFFT_TDSE_cel}\\
  \bigl[T_\text{N}+V_\text{N}(R)\bigr]\floqcnuc{R}&=E_\text{N}\floqcnuc{R}, \label{eq:HFFT_TDSE_cnuc}
\end{align}
with $E=\ee+\en$.
For a given $E_\text{e}$, Eq.~\eqref{eq:HFFT_TDSE_cel} is solved for each internuclear distance $R$ to obtain $\floqcel{x;R}$. These approximate continuum solutions were used successfully in Ref.~\cite{Madsen12}.

We find the solutions to Eq.~\eqref{eq:HFFT_TDSE_cel} numerically as follows. Starting near the origin, we impose the parity conditions $\floqcel{-\delta x;R}=(-1)^{P}\floqcel{\delta x;R}$, with $\delta x$ being the integration step size. We then apply the Numerov algorithm to numerically integrate Eq.~\eqref{eq:HFFT_TDSE_cel} outwards. The potential satisfies 
\begin{equation}
  V_\text{eN}(x,R) \rightarrow -\frac{2}{\abs{x}}, \quad \text{for} \; \abs{x} \gg R/2,
\end{equation}
which implies that the energy-normalized continuum solution has the asymptotic behavior
\begin{equation}
  \begin{aligned}
    \floqcel{x;R}
    \rightarrow 
    &\sqrt{\frac{\mu}{\pi p}} \bigl[ \coulombr\cos(\delta_P)\bigr.\\
    &\bigl. +\coulombi\sin(\delta_P) \bigr],
  \label{eq:HFFT_cel_asymp}
  \end{aligned}
\end{equation}
where $p=\sqrt{2\mu \ee}$, $\delta_P$ is the phaseshift, and $\coulombr$ and $\coulombi$ are the regular and irregular Coulomb functions, respectively. The latter functions were obtained using the GNU Scientific Library.
By matching our numerical solutions to the asymptotic form in Eq.~\eqref{eq:HFFT_cel_asymp}, we can obtain the energy $\delta$-normalized states satisfying $\braket{\floqcel{R}}{\xi^{P}_{{E_\text{e}}'}\left(R\right)}=\delta\left(\ee-{E_\text{e}}'\right)$.

%\bibliography{ref_2014_05}

\end{document}